# PROGRAMMING REQUESTS/RESPONSES WITH GREATFREE IN THE CLOUD ENVIRONMENT

Bing Li

Department of Computer Science and Engineering, Xi'An Technological University, China
bing.li@asu.edu

## ABSTRACT

*Programming request with GreatFree is an efficient programming technique to implement distributed polling in the cloud computing environment. GreatFree is a distributed programming environment through which diverse distributed systems can be established through programming rather than configuring or scripting. GreatFree emphasizes the importance of programming since it offers developers the opportunities to leverage their distributed knowledge and programming skills. Additionally, programming is the unique way to construct creative, adaptive and flexible systems to accommodate various distributed computing environments. With the support of GreatFree code-level Distributed Infrastructure Patterns, Distributed Operation Patterns and APIs, the difficult procedure is accomplished in a programmable, rapid and highly-patterned manner, i.e., the programming behaviors are simplified as the repeatable operation of Copy-Paste-Replace. Since distributed polling is one of the fundamental techniques to construct distributed systems, GreatFree provides developers with relevant APIs and patterns to program requests/responses in the novel programming environment.*

## KEYWORDS

*Code-Level Design Patterns, Cloud Programming, Distributed Systems, Highly-Patterned Development Environment*

## 1. INTRODUCTION

Programming requests/responses with GreatFree is an efficient programming technique to implement distributed polling in the cloud computing environment. Distributed polling [1] is an indispensable technique to construct distributed systems. When doing that with GreatFree, the procedure becomes straightforward since developers are not required to take care of underlying tough techniques. Using the rich APIs and patterns of DIP (Distributed Infrastructure Patterns) and DOP (Distributed Operation Patterns) supported by GreatFree, the tough programming skills are turned into simplified behaviors, the operation of CPR (Copy-Paste-Replace).[1]

GreatFree is a software development environment to construct cloud computing systems in diverse distributed computing environments through programming rather than configuring or scripting. Programming is defined as the procedure to implement a practical software system with essential domain knowledge and required programming skills. In the case of cloud programming, it is necessary for developers to perceive distributed knowledge as well as corresponding programming expertise. Although the overhead is high, it is the unique way to construct creative, adaptive and flexible systems with high-quality for various distributed computing environments. To lower the burden to program distributed systems, GreatFree provides developers with various code-level distributed patterns (DIP and DOP) and rich APIs.

---

[1] The research is sponsored by the Ministry of Education, Shaanxi Province, China. The number of the funding is 14JK1358.

Different from the object-oriented ones [2] which are independent of computing environments and the other distributed ones [3] which should be implemented by developers for specific environments, the code-level patterns in GreatFree specify the generic, mature and executable programs in the relatively fixed forms suitable to most distributed computing environments. For that, developers' effort is lowered even though they implement a distributed system through programming rather than scripting and configuring.

To be fond of them, developers need to perform the system-level programming initially. That is, they determine the infrastructure of their distributed systems through choosing the most suitable GreatFree DIP initially with respect to their distributed knowledge. And then, they need to accommodate the chosen DIP if it does not fulfill some of the requirements of the particular computing environment exactly with GreatFree DOP and APIs. After that, the system-level programming is completed and it constructs a high quality system foundation for developers to perform the application-level programming using GreatFree DOP and APIs further.

When programming with the operation of CPR, two concepts, i.e., the class reference and the instance reference, need to be identified by developers. The same as most object-oriented programming approaches, GreatFree is implemented in the language of Java [4] and it supports object-oriented programming for sure. The class reference specifies the class that need to be created newly for a new feature. The instance reference specifies that class instances that need to be created newly for a new feature. In the case of programming requests/responses through CPR, one sample request/response should be chosen from DIP to follow. Those references can be retrieved according to the sample. Once if those references are available, developers can create new classes or instances respectively through straightforward operation of CPR in the corresponding patterns of DOP.

The paper is organized in the following sections. Section 1 gives a brief introduction to the technique of programming requests/responses with GreatFree. Section 2 presents the related work to implement distributed polling in distributed computing environments and makes a rough comparison. Section 3 talks about the procedures to program cloud systems with GreatFree. Section 4 explains the details to program requests/responses with GreatFree through one case. Section 5 talks about the evaluation environment of the programming technique and discusses the potential future work of the technique.

## 2. RELATED WORK

Since distributed polling is one of the fundamental functions for any cloud systems, it is required to implement it with efficient approaches. Nowadays, there are three categories of solutions to the issue, including the traditional languages solutions, the framework-based solutions [5] and the programming-oriented solutions [6][7].

### 2.1 Traditional Languages Based Solutions

To adapt to the requirements of distributed computing environments, many new APIs are proposed to assist developers to do that. Traditional languages are defined as those programming languages that are originally established on the synchronous and standalone computing environment. In the obsolete environment, programming languages aim to solve the problems which can be processed in a sequential fashion within a single computer. That happens since the underlying CPU is designed in such a way as well as the computing resources are limited in terms of the processing power and the smallness of main memory when those languages are proposed initially.

Nowadays it is common to design asynchronous and distributed computing systems rather than synchronous and standalone ones. For that, many new APIs and techniques have to be put forward to accommodate them to the changes afterward. Although those techniques are succeeded in various application domains, it is well known that it is difficult to learn, grasp and program with them. The primary reason is due to the fact that each of them is founded on a synchronous and standalone computing environment such that the main purpose of those techniques is to transform the synchronous and standalone programming mechanism to the asynchronous and distributed one. Unfortunately, the transformation is visible to developers, i.e., they are enforced to take care of the heavy overhead before implementing their final high level applications. Because of the difficulty, only a small portion of developers would like to work with those techniques. It is often heard that one developer knows the concept of threading [8] whereas she never programs with it.

As one of the fundamental techniques of distributed computing, distributed polling is not compatible with the synchronous and standalone language either. Various techniques have to be resolved by developers themselves, such as communication [9], serialization [9], concurrency [8] and scaling [10], and so forth. In addition, since it is scarce that traditional programming languages are chosen as the development approach for distributed systems, it becomes almost impossible to program requests/responses or distributed polling with them.

## 2.2 Frameworks Based Solutions

Because of the difficulty to program with traditional programming languages to implement distributed polling as well as distributed systems, many frameworks [5] are proposed to hide distributed computing environments from developers. In consequence, for the issue of distributed polling, developers have no idea whether it happens in a standalone computer or a distributed environment when working on those frameworks.

The solution aims to convert a physically asynchronous and distributed computing system to a logically synchronous and standalone one for developers on the application level. Unluckily, it consists of numerous mutations in a distributed environment. According to the property of high-level applications, distributed systems are categorized into chatting, e-commerce, video, storage, search, social networks, and etc. With respect to transmission protocols, they are roughly categorized into the messaging one as well as the streaming one. It is reasonable to classify distributed environments into the one for heavyweight data and the one for lightweight data. It is also possible to identify distributed systems with the employed distributed models, such as the client/server one or the peer-to-peer one. Another perspective to observe distributed systems is to investigate its topology, such as the centralized one as well as the decentralized one. The scale is also an important indicator to describe distributed circumstances, such as the small-scale one and the large-scale one. Some systems work within a stable computing environment whereas others are located in a churning one. It frequently happens over the Internet. Another case over the Internet is that the huge differences exist among conceptual computing nodes in terms of their computing capacities. For that, the environment is identified as the homogeneous one and the heterogeneous one. The most difficult system might be the one that is dominated by human beings other than computers only [11][12]. The one is named as the social computing system [11][12] instead of the machine based one. In practice, it is usual that the above characters coexist in one particular environment such that it results in a more complicated computing system.

Because of the complexity of distributed environments, it is impossible to propose a system that hides all of the underlying techniques such that those frameworks focus on one specific domain that is widely engaged in popular applications. Within the narrow environment, the framework handles limited underlying techniques to transform an asynchronous and distributed system to a synchronous and standalone one. Thus, developers can implement their final applications through configuring or scripting rapidly through specifying their high level application requirements only within those dedicated frameworks. Although the approach is efficient, developers lose the opportunity to exploit their knowledge and skills to establish a system that exactly matches their functional and non-functional requirements. When a small incompatible obstacle is detected in the environment, it is difficult for them to handle. As of the issue of distributed polling, it is virtualized into the technique of conceptually synchronous method invocation. Developers cannot be aware of any distributed issues when using the technique. Similarly, it brings advantages as well as disadvantages.

## 2.3 New Languages Based Solutions

In recent years, some new languages [6][7] were proposed to adapt to the varieties of distributed computing environments. They attempt to provide developers with a new programming methodology that is asynchronous and distributed by nature. The philosophy is absolutely different from that of traditional ones, which are synchronous and standalone by default. When programming with those languages, the code is executed asynchronously without any explicit effort. That is, one subprogram is called through asynchronous messaging instead of synchronous invocation. Such a fundamental programming model [6][7] is validate in both of the local and distributed environments without any revisions. Using those languages, distributed polling is completely a built-in feature such that developers are not required to program it.

With such languages, developers are able to program distributed systems rather than work on one specific circumstance since the languages strive to be generic in the domain of distributed computing. Although it obtains the rapid development advantages, the technology of the new languages introduces some disadvantages as well. First of all, one of the potential problems of such a solution is that it forces developers to isolate from traditional programming experiences and work in a new programming context which updates the language expressions as well as the primary methodologies. Developers are enforced to think in a new asynchronous manner defined by those languages.

Additionally, the pursuit of the new languages is identical to that of the framework-based solutions. Both of them intend to hide developers from underlying techniques to raise the development speed and guarantee the quality of the final system. Different from the approach to provide application-specific script languages, the new languages claim that they support a concurrent and distributed programming model, such as Actor [6][7], which does not exist in traditional languages by default. It succeeds in speeding up the development procedure, but it fails in losing the possibility to open for changes. For example, Gossip [6] is employed as the protocol for Akka [6] to maintain the distributed nodes. In many cases of distributed environments, it is more practical to allow designers to deal with the issue since each environment usually needs to have its own appropriate algorithms for management.

More critical, the same as that of traditional programming ones, the methodology of the new languages assumes that the world can always be abstracted into a universal fundamental model. After the model is available, any scenarios in the computing environment can be programmed through a uniform approach. Unfortunately, when being confronted with the complexity of the Internet-based distributed environment, the assumption can hardly be correct all the time. For example, it is required for a large-scale and social-oriented system [11][12] to design a dedicated routing algorithm [13] to manage the overall topology and even states of each node. The algorithm should deal with the heterogeneous computing resources in a global scale though involving social theories [11][12]. Thus, the Akka built-in registry service has to be abandoned. In consequence, the high-level multicasting needs to be redesigned since it relies on the dedicated routing algorithm. Once if it happens, it represents the overall infrastructure of Akka is overwhelmed such that the Actor model is useless.

## 2.4 Comparisons

Different from all of the existing solutions, GreatFree proposes a series of patterns and APIs [14] to assist developers to come up with distributed systems. First of all, those APIs resolve most fundamental distributed problems that exist in diverse environments such that developers do not need to implement them with traditional programming languages. In addition, with the support of code-level patterns, including DIP and DOP, the programming procedure is conducted in a highly-patterned development environment to reach high efficiency and productivity. Finally, GreatFree does not believe there exists an abstract universal model in various distributed computing environments, especially over the Internet. To provide developers with a convenient programming environment, the APIs and patterns are always the building blocks adaptive to diverse cases. If those APIs and patterns are designed in a reasonable way, developers can compose them to any distributed systems with simplified behaviors according to distributed knowledge and programming skills arbitrarily.

## 3. PROGRAMMING REQUESTS/RESPONSES WITH GREATFREE

Programming requests/responses with GreatFree is a fundamental technique to implement the request-response-based remote interaction, i.e., distributed polling, between two remote computers within a distributed computing environment. Such an interaction is indispensable to construct any types of distributed systems, including the current popular one, i.e., the cloud.

### 3.1 The Philosophy

Different from traditional approaches, GreatFree provides developers with a highly-patterned environment to program requests/responses. First of all, developers implement requests/responses rapidly with GreatFree APIs [5] and code-level design patterns [5] without taking care of underlying techniques. During the procedure, they just need to perform simplified behaviors, i.e., the operation of CPR. In addition, with GreatFree, developers do not work in a virtualized standalone environment in which underlying techniques are invisible. Instead, at least, they should be aware of the basics of the computing environment. It is preferred that developers have sufficient knowledge about the Internet-based distributed environment in order to propose reasonable solutions. Meanwhile, they need to keep their programming skills when utilizing GreatFree APIs and patterns. The advantages of such a methodology not only guarantee the high efficiency of programming but also provide developers with the sufficient flexibility to deal with various distributed problems with their proficiency.

### 3.2 The Procedure of GreatFree Cloud Programming

The procedure of GreatFree programming is divided into two steps, the system-level phase and the application-level phase. The first one constructs the underlying foundation of the overall system that is suitable to one particular distributed environment. In contrast, the second one implements the upper application over the foundation established by the first phase.

#### 3.2.1 The System-Level Programming

When confronting with the problem to program within a distributed environment, first of all, developers need to consider which DIP (Distributed Infrastructure Patterns) is suitable to the domain. The DIP of GreatFree covers the most common cases of distributed computing environments in the real world. If the chosen DIP fits perfectly, developers can start to implement their upper level applications directly with GreatFree DOP and APIs. Then the system-level phase is terminated.

However, because of the rich mutations of distributed computing environments over the Internet, it is possible that the chosen DIP fulfills partial requirements only. In the case, developers have to expand the phase of the system-level programming in order to update the chosen DIP to accommodate the requirements in the particular environment using GreatFree DOP and APIs.

One extreme case is that no any DIP is suitable to the requirements of one particular distributed computing environment. If so, the development effort must be raised to a large extent. It needs to construct the particular infrastructure using GreatFree DOP and APIs heavily.

Only after the updated DIP meets the requirements of the specific circumstance, the system-level programming is accomplished and it is time to implement the applications upon it.

### 3.2.2 The Application-Level Programming

After the first phase is performed, a high-quality system foundation is constructed and it fits in the current distributed environment in a high degree. Thereafter, developers can program their upper applications. During the step, DOP and APIs are employed further to reach their final goals on the high level.

## 3.3 The Steps to Program Requests/Responses

As one of mandatory techniques to construct distributed systems, programming requests/responses with GreatFree is performed in a highly-patterned development environment. During the procedure, developers work with a limited number of GreatFree APIs and DOP only. More important, with the support of the highly-patterned environment, programming behaviors are simplified as the repeatable operations, i.e., CPR, through identifying and following class references and instance references. It represents that developers can implement a pair of request/response even though they have no idea about GreatFree methodology in the extreme case.

### 3.3.1 Choosing a Pair of Sample Request/Response

As a highly-patterned development environment, there are rich code-level patterns of DIP, DOP and APIs supported by GreatFree. It indicates that developers are able to follow existing code to implement their own systems. To program requests/responses, developers can choose the counterpart code from the environment in each step.

The first step is to choose one pair of request/response from the DIP developers work on. Since a DIP is a code-level system designed for one particular distributed environment and the technique of distributed polling is indispensable in any circumstances, it is convenient to locate one sample from the DIP. In addition, to help developers learn the development environment, some practical samples are implemented as open source. Therefore, it is convenient to choose one pair of sample request/response on the code-level to start up the programming.

### 3.3.2 Creating the Request/Response

To ease the design, after the sample request/response is chosen, developers do not need to program the request/response from scratch. Instead, since each pair of request/response encloses the data to be sent, developers just need to remain the request/response pattern of SM (Server Message) [14] and replace the existing data with the data they attempt to send. If the data is primitive, the replacement is performed straightforward. If the data is self-defined, i.e., a class, it is necessary to serialize it since it is required to be transmitted within a distributed environment.

### 3.3.3 Retrieving Class References

The concept of the class references is defined as one independent class that references one distributed message in GreatFree development environment. Meanwhile, the class is programmed particularly as new ones for a distributed computing environment in a predefined code-level pattern. The class references can be easily retrieved for developers to follow through CPR. In the case of programming requests/responses, the class references exist at the server side only, including a thread that processes the incoming request concurrently and a thread creator that is injected into the underlying thread pool to create new instances of the thread in case that the existing ones cannot deal with the volume of incoming requests. It is convenient for developers to retrieve the class references, i.e., the thread and the thread creator, with any programming tools.

### 3.3.4 Retrieving Instance References

The concept of instance references specifies the newly created class instances in the existing components, which are essential modules in DIP to keep the particular infrastructure to work in a high-quality manner. Similar to class references, they are written in the code-level highly-patterned form such that it is convenient for developers to follow through CPR. In the case of programming requests/responses, the instance references exist at both of the server side and the client side. At the server side, the instance reference is written in the pattern of RD (Request Dispatcher) [14] in the component of the server dispatcher, which is structured in the pattern, SD (Server Dispatcher) [14], as well. At the client side, the instance reference is located in the pattern of RR (Remote Reader) [14]. The pattern exists in the component of the client reader, which is placed in the pattern of CR (Client Reader) [14] .

One special case is that one instance reference is not declared explicitly with an instance name in the sample code, but it is referenced by type casting, the new keyword and so forth. It happens frequently to simplify the syntax. The one is called the implicit instance reference. In contrast, the one with a declared name is called the explicit instance reference. Whichever the types of instance references are, the programming behavior is always the operation of CPR. Usually, an implicit instance reference is more straightforward to be followed since it is referenced independently in the form of the class.

### 3.3.5 Creating the Thread and Its Creator

The thread is responsible for processing incoming requests concurrently at the server side. In GreatFree, each type of requests/responses has one dedicated thread. If programming from scratch, it is a time-consuming and error-prone task. With the support of GreatFree, developers is only required to follow the class references retrieved in the fourth step through CPR. To do that, developers create a new class named in a proper convention at first. Then, they should keep the pattern through the operations of copying/pasting from the class reference. At least, they have to replace the sample request/response and the sample processing code with their own.

Another class reference is the thread creator that should accompany with the newly created thread. The reference is created newly as well. Fortunately, the procedure is straightforward through CPR, i.e., keeping the pattern and replacing the sample request/response and the sample thread.

### 3.3.6 Updating the Server Dispatcher

After the thread and its creator are created, they should be injected into the server side to process incoming requests. To achieve the goal in GreatFree, developers are required to perform the operation of CPR on the instance reference in the pattern of RD (Request Dispatcher) and SE (Server Dispatcher) at the server side. The instance reference contains the samples of the request/response, the thread and the thread creator. Thus, to do that through CPR, at first, developers ought to create a new instance for the new request/response following the instance reference and replace its components, the samples of the request/response, the thread ant thread creator with those new classes which are created following the corresponding class references in relevant patterns. Then, for each place where the instance reference is located, perform the operation of CPR again in the patterns of RD and SD.

### 3.3.7 Updating the Client Reader

Compared with the server side, the client side is simple since it is not necessary to take into account the issue of processing high volume of incoming requests. Instead, the unique task of the client is responsible for sending the request to the server and wait until the response from the server is received. Thus, at the client side, the instance reference contains the sample request/response only. Similar to what is performed on the server side, developers just need to create a new instance containing the new request/response following the instance reference and then perform the operation of CPR at each location of the instance reference again in the pattern of CR (Client Reader).

## 4. CASE STUDY

To program requests/responses, it is suggested to follow the order, i.e., the request/response, the server side and then the client side. No matter which step to work on, developers' programming behaviors are simplified as CPR. The steps are presented with a simple example, i.e., sending one request from one client to one server to retrieve certain information and holding on at the client until the response enclosing the retrieved information is received.

### 4.1 The Request/Response

To start the programming, one pair of sample request/response should be located from the DIP or any other sample projects in open source. For example, one sample request, WeatherRequest.java/WeatherResponse.java, is chosen and shown in Listing 1 and Listing 2, respectively, from the relevant Java package of the DIP. This is the request/response that developers can follow to design their own through CPR.

```
1    public class WeatherRequest extends ServerMessage
2    {
3        public WeatherRequest()
4        {
5            super(MessageType.WEATHER_REQUEST);
6        }
7    }
```
Listing 1. The Code of WeatherRequest.java

```
1    public class WeatherResponse extends ServerMessage
2    {
3        // Declare the instance of Weather. 02/15/2016, Bing Li
4        private Weather weather;
5    
6        public WeatherResponse(Weather weather)
7        {
8            super(MessageType.WEATHER_RESPONSE);
9            this.weather = weather;
10       }
11   
12       public Weather getWeather()
13       {
14           return this.weather;
15       }
16   }
```
Listing 2. The Code of WeatherResponse.java

After that, the request/response, i.e., TestRequest.java/TestResponse.java, are listed as shown in Listing 3 and Listing 4, respectively.

```
1    public class TestRequest extends ServerMessage
2    {
3        private String request;
4    
5        public TestRequest(String request)
```

```
6       {
7           super(MessageType.TEST_REQUEST);
8           this.request = request;
9       }
10
11      public String getRequest()
12      {
13          return this.request;
14      }
15  }
```

Listing 3. The Code of TestRequest.java

```
1   public class TestResponse extends ServerMessage
2   {
3       private String response;
4
5       public TestResponse(String response)
6       {
7           super(MessageType.TEST_RESPONSE);
8           this.response = response;
9       }
10
11      public String getResponse()
12      {
13          return this.response;
14      }
15  }
```

Listing 4. The Code of TestResponse.java

Comparing the structure of WeatherRequest/WeatherResponse with that of TestRequest/TestResponse, they are identical except the data to be exchanged. The WeatherRequest encloses nothing whereas the TestRequest contains only one String, request. More important, both of them are written in the pattern of SM (Server Message). Correspondingly, the WeatherResponse encloses the self-defined data type, Weather, whereas the TestResponse just includes the contrived data in the type of String. Because of the identical structure of requests/responses in GreatFree, it is convenient to create new ones through the simplified behaviors, i.e., CPR. That represents that GreatFree is a highly-patterned programming environment. Thus, CPR can be performed in GreatFree.

### 4.2 The Class References to the Sample Request/Response

To continue programming requests/responses using the approach of CPR, it is necessary to find the additional samples to follow. The sample code is easily retrieved within the DIP source code according to the request. If an appropriate IDE (Integrated Development Environment) [15] is used, the procedure becomes more convenient. For example, if Eclipse [15] is the chosen IDE, developers can retrieve the corresponding sample by searching the references to WeatherRequest, as shown in Figure 1. Through the approach, developers can locate all of the relevant code of the sample request. Each of the code should be followed through CPR at the server side as well as the client side. Figure 2 presents the references to WeatherResponse.

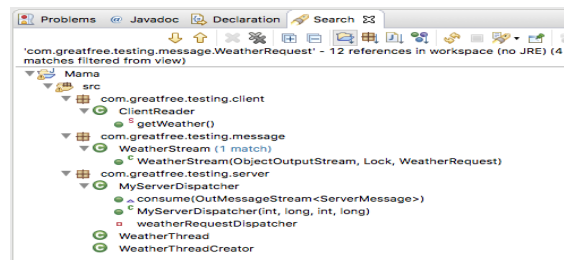

Figure 1. The References to the Sample Request, WeatherRequest

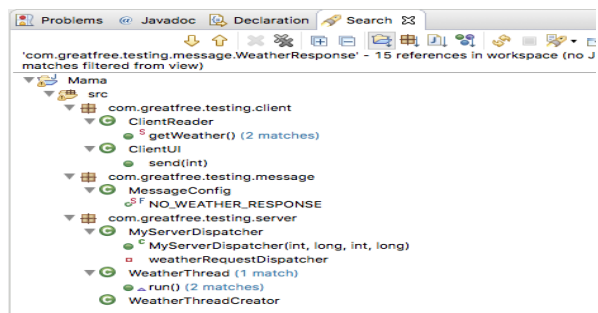

Figure 2. The References of the Sample Response, WeatherResponse

All of the references retrieved in this case are summarized in Table 1. All of them are the sample code developers need to follow for the new created request/response, TestRequest/TestResponse. The references to the messages exist on both sides, the client and the server, since the request and the response are the messages that are exchanged between the client and the server.

Table 1. The References to the Sample Request/Response, WeatherRequest/WeatherResponse

| Class | Explanation | Side |
| --- | --- | --- |
| ClientReader | The client side that sends requests to the server side and waits until a response is received | Client |
| ClientUI | The UI presenter at the client side | Client |
| MessageConfig | The configurations of messages | N/A |
| WeatherStream | The output stream that sends the response to the client | N/A |
| MyServerDispacher | The dispatcher that is responsible for dispatching received messages to corresponding threads such that those messages are processed concurrently | Server |
| WeatherThread | The thread that deals with the request of WeatherRequest | Server |
| WeatherThreadCreator | The thread that creates a new instance of WeatherThread in case that existing instances are too busy | Server |

The next task is to determine the new classes to be created for TeststRequest/TestResponse. Table 1 lists seven classes, i.e., ClientReader, ClientUI, MessageConfig, WeatherStream, MyServerDispacher, WeatherThread and WeatherThreadCreator. Although the names of them are not necessarily given in any conventions, it is highly suggested to append the suffixes, i.e., -Reader, -UI, -Config, -Stream, -ServerDispatcher, -Thread and -ThreadCreator, at the end of each of them. On the other hand, the prefix of each class reference is preferred to exhibit the request/response. Hence, according to the names of the classes, it is convenient to infer that the three ones, WeatherStream, WeatherThread and WeatherThreadCreator, are designed particularly for the corresponding request/response whereas the other four are existing ones for any messages. The ones, which are designed particularly for WeatherRequest/WeatherResponse, are called class references. Afterward, according to the class references to the sample request/response, developers are able to create new classes for TestRequest/TestResponse, i.e., TestStream, TestRequestThread and TestRequestThreadCreator, which are named in the same convention as their respective counterparts, i.e., WeatherStream, WeatherThread and WeatherThreadCreator. Table 2 and Table 3 list the class references to WeatherRequest/WeatherResponse and their counterparts, which should be programmed newly by developers.

Table 2. The Class References to WeatherRequest and Their Counterparts for TestRequest

| Type | Class References to WeatherRequest | Class References to TestRequest |
| --- | --- | --- |
| Request | WeatherRequest | TestRequest |
| Stream | WeatherStream | TestStream |
| Thread | WeatherThread | TestRequestThread |
| Thread Creator | WeatherThreadCreator | TestRequestThreadCreator |

Table 3. The Class References to WeatherResponse and Their Counterparts for TestResponse

| Type | Class References to WeatherResponse | Class References to TestResponse |
| --- | --- | --- |
| Response | WeatherResponse | TestResponse |
| Thread | WeatherThread | TestRequestThread |
| Thread Creator | WeatherThreadCreator | TestRequestThreadCreator |

In short, Table 1 helps developers to locate all of the additional sample code to follow. When programming new classes further, developers should CPR each of the class reference counterparts based on Table 2 and Table 3.

### 4.3 The Instance References to the Sample Request/Response

Besides class references to the sample request/response, it is required to follow instance references. In accordance with Table 1, there are four existing classes, ClientReader, ClientUI, MessageConfig and MyServerDispatcher, which are not designed particularly for the request/response of WeatherRequest/WeatherResponse. Instead, they are used to process any requests/responses at the client side and the server side, respectively. Inside the four classes, some class instances reference the WeatherRequest/WeatherResponse. Table 4 lists the instance references to WeatherRequest/WeatherResponse in the four classes.

Table 4. The Instance References to WeatherRequest/WeatherResponse and Their CounterParts to TestRequest/TestResponse

| Class Instance | Type | Location | Counterpart |
| --- | --- | --- | --- |
| Implicit reference to WeatherRequest | WeatherRequest | ClientReader | TestRequest |
| Implicit reference to WeatherResponse | WeatherResponse | ClientReader | TestResponse |
| weatherResponse | WeatherResponse | ClientUI | testResponse |
| NO_WEATHER_RESPONSE | WeatherResponse | MessageConfig | NO_TEST_RESPONSE |
| weatherRequestDispatcher | RequestDispatcher<WeatherRequest, WeatherStream, WeatherResponse, WeatherThread, WeatherThreadCreator> | MyServerDispatcher | testRequestDispatcher |

With respect to Table 4, developers need to program the instance references, including those implicit instance references in ClientReader, testResponse in ClientUI, NO_TEST_RESPONSE in MessageConfig and testRequestDispatcher in MyServerDispatcher, following the instance references with the approach of CPR.

### 4.4 The Server Side

For a pair of request/response, most effort is usually spent on the server side since it has to deal with potentially high volume of accessing load, including incoming requests.

#### 4.4.1 The Stream

According to Table 2, WeatherRequest has one class reference, WeatherStream. Besides creating messages themselves, one additional effort is required when programming requests/responses, compared with programming requests/responses. It is required to create a new class, TestStream as shown in Listing 6 following its counterpart, WeatherStream as shown in Listing 5, to complete the message programming. TestStream is used to send the response to the client at the server side.

```
1    public class WeatherStream extends OutMessageStream<WeatherRequest>
2    {
3        public WeatherStream(ObjectOutputStream out, Lock lock, WeatherRequest message)
4        {
5            super(out, lock, message);
6        }
7    }
```
Listing 5. The code of WeatherStream.java

The structure of WeatherStream is simple and it is straightforward to follow it through CPR. Thus, the counterpart, TestStream, is created as shown in Listing 6.

```
1    public class TestStream extends OutMessageStream<TestRequest>
2    {
3        public TestStream(ObjectOutputStream out, Lock lock, TestRequest message)
4        {
5            super(out, lock, message);
6        }
7    }
```
Listing 6. The code of TestStream.java

#### 4.4.2 The Thread

To process requests concurrently, a thread, e.g., TestRequestThread, should be designed for that. Its counterpart, WeatherThread, which is one class reference of WeatherRequest, is shown in Listing 7. It needs to indicate that one important pattern, RDWC (Request Double-While-Concurrency) [14], is used in the thread. This is one of the DOP in GreatFree. When programming a thread and attempting to keep it under control by a pooling algorithm [14], it is required to employ the pattern.

```
1    public class WeatherThread extends RequestQueue<WeatherRequest, WeatherStream, WeatherResponse>
2    {
3        public WeatherThread(int maxTaskSize)
4        {
5            super(maxTaskSize);
6        }
7
8        public void run()
9        {
10            WeatherStream request;
11           WeatherResponse response;
12           while (!this.isShutdown())
13           {
14               // The loop detects whether the queue is empty or not. 02/15/2016, Bing Li
15               while (!this.isEmpty())
16               {
17                   // Dequeue a request. 02/15/2016, Bing Li
18                   request = this.getRequest();
19                   // Initialize an instance of WeatherResponse. 02/15/2016, Bing Li
```

```
20                          response = new WeatherResponse(WeatherDB.SERVER().getWeather());
21                          try
22                          {
23                              // Respond the response to the remote client. 02/15/2016, Bing Li
24                              this.respond(request.getOutStream(), request.getLock(), response);
25                          }
26                          catch (IOException e)
27                          {
28                              e.printStackTrace();
29                          }
30                          // Dispose the messages after the responding is performed. 02/15/2016, Bing Li
31                          this.disposeMessage(request, response);
32                      }
33                      try
34                      {
35                          this.holdOn(ServerConfig.REQUEST_THREAD_WAIT_TIME);
36                      }
37                      catch (InterruptedException e)
38                      {
39                          e.printStackTrace();
40                      }
41                  }
42              }
43          }
```

Listing 7. The code of WeatherThread.java

After CPR, the thread of TestRequestThread is shown in Listing 8.

```
1    public class TestRequestThread extends RequestQueue<TestRequest, TestStream, TestResponse>
2    {
3        public TestRequestThread(int maxTaskSize)
4        {
5            super(maxTaskSize);
6        }
7
8        public void run()
9        {
10           TestStream request;
11           TestResponse response;
12           // The loop detects whether the queue is empty or not. 02/15/2016, Bing Li
13           while (!this.isShutdown())
14           {
15               // The loop detects whether the queue is empty or not. 02/15/2016, Bing Li
16               while (!this.isEmpty())
17               {
18                   // Dequeue a request. 02/15/2016, Bing Li
19                   request = this.getRequest();
20                   // Initialize an instance of WeatherResponse. 02/15/2016, Bing Li
21                   response = new TestResponse("response");
22                   try
23                   {
24                       // Respond the response to the remote client. 02/15/2016, Bing Li
25                       this.respond(request.getOutStream(), request.getLock(), response);
26                   }
27                   catch (IOException e)
28                   {
29                       e.printStackTrace();
30                   }
31                   // Dispose the messages after the responding is performed. 02/15/2016, Bing Li
32                   this.disposeMessage(request, response);
33               }
34               try
35               {
36                   this.holdOn(ServerConfig.REQUEST_THREAD_WAIT_TIME);
37               }
38               catch (InterruptedException e)
39               {
40                   e.printStackTrace();
41               }
42           }
43       }
44   }
```

Listing 8. The code of TestRequestThread.java

Comparing the code of TestRequestThread with its counterpart, WeatherThread, the structure, i.e., the pattern of RDWC, is identical. The distinctions occur in the request/response types and the operations to process the messages. It proves further that GreatFree is a highly-patterned programming environment.

### 4.5.2 The Thread Creator

The thread creator is employed by the underlying thread pool to create new instances of the thread to process requests when existing ones can hardly deal with the accessing load. As the class reference to WeatherRequest as well as WeatherResponse, the sample code of WeatherThreadCreator is shown in Listing 9.

```
1    public class WeatherThreadCreator implements RequestThreadCreatable<WeatherRequest,
2        WeatherStream, WeatherResponse, WeatherThread>
3    {
4        @Override
5        public WeatherThread createRequestThreadInstance(int taskSize)
6        {
7            return new WeatherThread(taskSize);
8        }
9    }
```

Listing 9. The code of WeatherThreadCreator.java

The corresponding code to be programmed is named as TestRequestThreadCreator. To program it with CPR, the structure remains whereas the request, the stream, the response and the thread should be replaced with TestRequest, TestStream, TestResponse and TestRequestThread. The code is shown in Listing 9.

```
1    public class TestRequestThreadCreator implements RequestThreadCreatable<TestRequest,
2        TestStream, TestResponse, TestRequestThread>
3    {
4        @Override
5        public TestRequestThread createRequestThreadInstance(int taskSize)
6        {
7            return new TestRequestThread(taskSize);
8        }
9    }
```

Listing 10. The code of TestRequestThreadCreator.java

### 4.4.3 The Server Dispatcher

The server dispatcher is the most important component for a distributed node as a server. It implements the primary underlying techniques as an independent computing unit in a distributed environment. After the request/response and the thread are available, it needs to embed them into the server dispatcher to finish the programming on the server side.

Different from the previous programming to create new classes following class references to one pair of request/response, developers are required to make revisions to the existing server dispatcher in the case through following instance references to the sample request/response. According to Table 4, as the instance reference to WeatherRequest/WeatherResponse in MyServerDispatcher, weatherRequestDispatcher, is the sample code to be followed by its counterpart. Developers just need to create a new instance, testRequestDispatcher, in the server dispatcher. Fortunately, the entire procedure is still straightforward to be performed in a highly-patterned environment through CPR.

Listing 11 presents the code of the server dispatcher, MyServerDispatcher, which derives the API, ServerMessageDispatcher. To save space, only the relevant instance references are shown in the list.

```
1    public class MyServerDispatcher extends ServerMessageDispatcher<ServerMessage>
2    {
3        ……
6        private RequestDispatcher<WeatherRequest, WeatherStream, WeatherResponse, WeatherThread,
7            WeatherThreadCreator> weatherRequestDispatcher;
8        ……
9        public MyServerDispatcher(int threadPoolSize, long threadKeepAliveTime, int schedulerPoolSize,
10           long schedulerKeepAliveTime)
11       {
12           super(threadPoolSize, threadKeepAliveTime, schedulerPoolSize, schedulerKeepAliveTime);
13           ……
14
15           this.weatherRequestDispatcher = new RequestDispatcher.RequestDispatcherBuilder
16               <WeatherRequest, WeatherStream, WeatherResponse, WeatherThread, WeatherThreadCreator>()
17               .poolSize(ServerConfig.REQUEST_DISPATCHER_POOL_SIZE)
```

```
18              .keepAliveTime(ServerConfig.REQUEST_DISPATCHER_THREAD_ALIVE_TIME)
19              .threadCreator(new WeatherThreadCreator())
20              .maxTaskSize(ServerConfig.MAX_REQUEST_TASK_SIZE)
21              .dispatcherWaitTime(ServerConfig.REQUEST_DISPATCHER_WAIT_TIME)
22              .waitRound(ServerConfig.REQUEST_DISPATCHER_WAIT_ROUND)
23              .idleCheckDelay(ServerConfig.REQUEST_DISPATCHER_IDLE_CHECK_DELAY)
24              .idleCheckPeriod(ServerConfig.REQUEST_DISPATCHER_IDLE_CHECK_PERIOD)
25              .scheduler(super.getSchedulerPool())
26              .build();
27          ……
28      }
29
30      // Shut down the server message dispatcher. 09/20/2014, Bing Li
31      public void shutdown() throws InterruptedException
32      {
33          ……
34          // Dispose the weather request dispatcher. 02/15/2016, Bing Li
35          this.weatherRequestDispatcher.dispose();
36          ……
37          // Shutdown the derived server dispatcher. 11/04/2014, Bing Li
38          super.shutdown();
39      }
40
41      // Process the available messages in a concurrent way. 09/20/2014, Bing Li
42      public void consume(OutMessageStream<ServerMessage> message)
43      {
44          // Check the types of received messages. 11/09/2014, Bing Li
45          switch (message.getMessage().getType())
46          {
47              ……
48              // If the message is the one of weather requests. 11/09/2014, Bing Li
49              case MessageType.WEATHER_REQUEST:
50                  // Check whether the weather request dispatcher is ready. 02/15/2016, Bing Li
51                  if (!this.weatherRequestDispatcher.isReady())
52                  {
53                      // Execute the weather request dispatcher concurrently. 02/15/2016, Bing Li
54                      super.execute(this.weatherRequestDispatcher);
55                  }
56                  // Enqueue the instance of WeatherRequest into the dispatcher for
57                  // concurrent responding. 02/15/2016, Bing Li
58                  this.weatherRequestDispatcher.enqueue(new WeatherStream(message.getOutStream(),
59                      message.getLock(), (WeatherRequest)message.getMessage()));
60                  break;
61              ……
62          }
63      }
64  }
```

Listing 11. The code of MyServerDispatcher.java

According to the code in Listing 11, the instance reference to WeatherRequest/WeatherResponse, weatherRequestDispatcher, is written in the syntax of Generics [16]. It encloses five components, including WeatherRequest, WeatherStream, WeatherResponse, SetWeatherThread and SetWeatherThreadCreator, all of which are the class references to WeatherRequest/WeatherResponse. Thus, when programming its counterpart, testRequestDispatcher, what developers should do first is to copy the instance reference and replace the instance references as well as the class references with their counterparts, respectively. It needs to repeat the operations for all of the lines where the instance reference emerges. The entire CPR procedure is performed in a straightforward manner.

During the steps, all of the previous followed sample code, i.e., class references and instance references, coexist in the server dispatcher. They form some of the important DOP patterns, i.e., the Server Dispatcher (SD) [14] and the Request Dispatcher (RD) [14]. That is why developers are able to update the code with simplified behaviors, i.e., CPR.

After all of the above steps, the programming on the server side is done and the code is updated as shown in Listing 12.

```
1   public class MyServerDispatcher extends ServerMessageDispatcher<ServerMessage>
2   {
3       ……
4       private RequestDispatcher<WeatherRequest, WeatherStream, WeatherResponse, WeatherThread,
5           WeatherThreadCreator> weatherRequestDispatcher;
6
7       private RequestDispatcher<TestRequest, TestStream, TestResponse, TestRequestThread,
8           TestRequestThreadCreator> testRequestDispatcher;
9       ……
```

```java
10      public MyServerDispatcher(int threadPoolSize, long threadKeepAliveTime, int schedulerPoolSize,
11          long schedulerKeepAliveTime)
12      {
13          super(threadPoolSize, threadKeepAliveTime, schedulerPoolSize, schedulerKeepAliveTime);
14          ……
15          this.weatherRequestDispatcher = new RequestDispatcher.RequestDispatcherBuilder
16              <WeatherRequest, WeatherStream, WeatherResponse, WeatherThread, WeatherThreadCreator>()
17                  .poolSize(ServerConfig.REQUEST_DISPATCHER_POOL_SIZE)
18                  .keepAliveTime(ServerConfig.REQUEST_DISPATCHER_THREAD_ALIVE_TIME)
19                  .threadCreator(new WeatherThreadCreator())
20                  .maxTaskSize(ServerConfig.MAX_REQUEST_TASK_SIZE)
21                  .dispatcherWaitTime(ServerConfig.REQUEST_DISPATCHER_WAIT_TIME)
22                  .waitRound(ServerConfig.REQUEST_DISPATCHER_WAIT_ROUND)
23                  .idleCheckDelay(ServerConfig.REQUEST_DISPATCHER_IDLE_CHECK_DELAY)
24                  .idleCheckPeriod(ServerConfig.REQUEST_DISPATCHER_IDLE_CHECK_PERIOD)
25                  .scheduler(super.getSchedulerPool())
26                  .build();
27
28          this.testRequestDispatcher = new RequestDispatcher.RequestDispatcherBuilder
29              <TestRequest, TestStream, TestResponse, TestRequestThread, TestRequestThreadCreator>()
30                  .poolSize(ServerConfig.REQUEST_DISPATCHER_POOL_SIZE)
31                  .keepAliveTime(ServerConfig.REQUEST_DISPATCHER_THREAD_ALIVE_TIME)
32                  .threadCreator(new TestRequestThreadCreator())
33                  .maxTaskSize(ServerConfig.MAX_REQUEST_TASK_SIZE)
34                  .dispatcherWaitTime(ServerConfig.REQUEST_DISPATCHER_WAIT_TIME)
35                  .waitRound(ServerConfig.REQUEST_DISPATCHER_WAIT_ROUND)
36                  .idleCheckDelay(ServerConfig.REQUEST_DISPATCHER_IDLE_CHECK_DELAY)
37                  .idleCheckPeriod(ServerConfig.REQUEST_DISPATCHER_IDLE_CHECK_PERIOD)
38                  .scheduler(super.getSchedulerPool())
39                  .build();
40          ……
41      }
42
43      public void shutdown() throws InterruptedException
44      {
45          ……
46          // Dispose the weather request dispatcher. 02/15/2016, Bing Li
47          this.weatherRequestDispatcher.dispose();
48          this.testRequestDispatcher.dispose();
49          ……
50          // Shutdown the derived server dispatcher. 11/04/2014, Bing Li
51          super.shutdown();
52      }
53
54      // Process the available messages in a concurrent way. 09/20/2014, Bing Li
55      public void consume(OutMessageStream<ServerMessage> message)
56      {
57          // Check the types of received messages. 11/09/2014, Bing Li
58          switch (message.getMessage().getType())
59          {
60              ……
61              // If the message is the one of weather requests. 11/09/2014, Bing Li
62              case MessageType.WEATHER_REQUEST:
63                  // Check whether the weather request dispatcher is ready. 02/15/2016, Bing Li
64                  if (!this.weatherRequestDispatcher.isReady())
65                  {
66                      // Execute the weather request dispatcher concurrently. 02/15/2016, Bing Li
67                      super.execute(this.weatherRequestDispatcher);
68                  }
69                  // Enqueue the instance of WeatherRequest into the dispatcher for
70                  // concurrent responding. 02/15/2016, Bing Li
71                  this.weatherRequestDispatcher.enqueue(new WeatherStream(message.getOutStream(),
72                      message.getLock(), (WeatherRequest)message.getMessage()));
73                  break;
74
75              case MessageType.TEST_REQUEST:
76                  System.out.println("TEST_REQUEST received @" + Calendar.getInstance().getTime());
77                  // Check whether the test request dispatcher is ready. 02/15/2016, Bing Li
78                  if (!this.testRequestDispatcher.isReady())
79                  {
80                      // Execute the test request dispatcher concurrently. 02/15/2016, Bing Li
81                      super.execute(this.testRequestDispatcher);
82                  }
83                  // Enqueue the instance of WeatherRequest into the dispatcher for
84                  // concurrent responding. 02/15/2016, Bing Li
85                  this.testRequestDispatcher.enqueue(new TestStream(message.getOutStream(),
86                      message.getLock(), (TestRequest)message.getMessage()));
87                  break;
88              ……
89          }
90      }
91  }
```

Listing 12. The code of MyServerDispatcher.java after CPR

## 4.5 The Client Side

The last step to program requests/responses is to update the instance reference at the client side. To do that, the programming effort is spent on two portions, i.e., following the implicit instance references and following the explicit instances.

**4.5.1 Following the Implicit Instance References**

According to Table 4, only implicit references exist in the code of ClientReader. To program the client side, it still needs to perform the operation of CPR on those implicit instance references.

Before making any updates, the code of ClientReader is shown in Listing 13. The same as MyServerDispatcher in Listing 11, only implicit instance references are presented and others are omitted to save space.

```
1    public class ClientReader
2    {
3        ……
4        public static WeatherResponse getWeather()
5        {
6            try
7            {
8                return (WeatherResponse)(RemoteReader.REMOTE().read(NodeID.DISTRIBUTED().getKey(),
9                    ServerConfig.SERVER_IP, ServerConfig.SERVER_PORT, new WeatherRequest()));
10           }
11           catch (ClassNotFoundException | RemoteReadException | IOException e)
12           {
13               e.printStackTrace();
14           }
15           return MessageConfig.NO_WEATHER_RESPONSE;
16       }
17       ……
18   }
```

Listing 13. The Code of ClientReader.java

To follow the implicit instance references, WeatherRequest/WeatherResponse, the operation is more convenient since those references are usually used for type casting or creating new instances independently such that the code is more simple. To program their counterparts, TestRequest/TestResponse, just copy the lines where the implicit instances emerge and replace each of them. During the procedure, the programming is performed within one DOP, RR (Remote Reader).

After that, the code is updated as shown in Listing 14.

```
1    public class ClientReader
2    {
3        ……
4        public static WeatherResponse getWeather()
5        {
6            try
7            {
8                return (WeatherResponse)(RemoteReader.REMOTE().read(NodeID.DISTRIBUTED().getKey(),
9                    ServerConfig.SERVER_IP, ServerConfig.SERVER_PORT, new WeatherRequest()));
10           }
11           catch (ClassNotFoundException | RemoteReadException | IOException e)
12           {
13               e.printStackTrace();
14           }
15           return MessageConfig.NO_WEATHER_RESPONSE;
16       }
17
18       public static TestResponse getResponse(String request)
19       {
20           try
21           {
22               return (TestResponse)(RemoteReader.REMOTE().read(NodeID.DISTRIBUTED().getKey(),
23                   ServerConfig.SERVER_IP, ServerConfig.SERVER_PORT, new TestRequest(request)));
24           }
25           catch (ClassNotFoundException | RemoteReadException | IOException e)
26           {
27               e.printStackTrace();
28           }
29           return MessageConfig.NO_TEST_RESPONSE;
30       }
31       ……
32   }
```

Listing 14. The Code of ClientReader After CPR

### 4.5.2 Following the Explicit Instance References

The last step is to program through following the explicit instance references. In the case, only one explicit instance reference to WeatherResponse exists in the code of ClientUI at the client side. Since it is necessary to display the response on the screen, it takes a little effort in the step. In a more general case which does not need to present the response, the step is ignored.

The code of ClientUI before programming is shown in Listing 15.

```
1     public class ClientUI
2     {
3         ……
4         public void send(int option)
5         {
6             ……
7             WeatherResponse weatherResponse;
8             Weather weather;
9             // Check the option to interact with the polling server. 09/21/2014, Bing Li
10            switch (option)
11            {
12                ……
13                // If the GET_WEATHER option is selected, send the request message to the
14                // remote server. 02/18/2016, Bing Li
15                case MenuOptions.GET_WEATHER:
16                    weatherResponse = ClientReader.getWeather();
17                    weather = weatherResponse.getWeather();
18                    System.out.println("Temperature: " + weather.getTemperature());
19                    System.out.println("Forcast: " + weather.getForcast());
20                    System.out.println("Rain: " + weather.isRain());
21                    System.out.println("How much rain: " + weather.getHowMuchRain());
22                    System.out.println("Time: " + weather.getTime());
23                    break;
24                ……
25            }
26        }
27    }
```

Listing 15. The Code of ClientUI.java

The procedure to follow the explicit instance reference to WeatherResponse at the client side is identical to that at the server side. After CPR, the update code of ClientUI.java is shown in Listing 16.

```
1     public class ClientUI
2     {
3         ……
4         public void send(int option)
5         {
6             ……
7             WeatherResponse weatherResponse;
8             Weather weather;
9             TestResponse testResponse;
10
11            // Check the option to interact with the polling server. 09/21/2014, Bing Li
12            switch (option)
13            {
14                ……
15                // If the GET_WEATHER option is selected, send the request message to the
16                // remote server. 02/18/2016, Bing Li
17                case MenuOptions.GET_WEATHER:
18                    weatherResponse = ClientReader.getWeather();
19                    weather = weatherResponse.getWeather();
20                    System.out.println("Temperature: " + weather.getTemperature());
21                    System.out.println("Forcast: " + weather.getForcast());
22                    System.out.println("Rain: " + weather.isRain());
23                    System.out.println("How much rain: " + weather.getHowMuchRain());
24                    System.out.println("Time: " + weather.getTime());
25                    break;
26
27                case MenuOptions.REQUEST_TEST:
28                    testResponse = (TestResponse)ClientReader.getResponse("request");
29                    System.out.println(testResponse.getResponse());
30                    break;
31                ……
32            }
33        }
34    }
```

Listing 16. The Code of ClientUI.java after CPR

## 4.6 Testing

Now it is time to run the code of distributed polling implemented by the technique of programming requests/responses with GreatFree. It is necessary to program some code that invokes the client reader. To save space, it is omitted in the paper. The detailed information can be found in the Chapter 4 of the book, Programming Clouds With GreatFree [14].

The server should be started up at first and then the client is executed. A new option, Request Test, is displayed in the menu. Just select the option, Then, a message, "TEST_REQUEST", is displayed on the server side (Figure 3). It represents that distributed polling is injected into the DIP successfully. At the client side, a "response" is received and display (Figure 4).

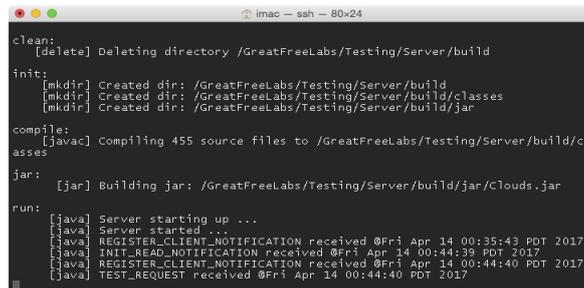

Figure 3. The message, "TEST_REQUEST", of the new request, TestRequest, is displayed on the server side

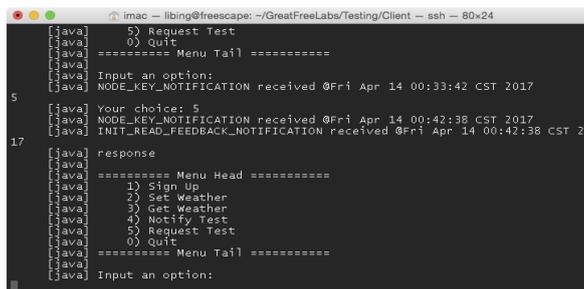

Figure 4. The client side after the new request is sent and the response is received

## 5. EVALUATION ANF FUTURE WORK

GreatFree cloud programming environment was implemented during the large research project, the New World Wide Web [13]. The project proposes many new fundamental algorithms to resolve. To implement them, it is impossible to employ any existing frameworks for distributed computing environments. Instead, traditional generic programming language, i.e., Java SE [4], has to be selected as the primary development approach. Because the system is huge, it has been taking a couple of years. All of the source code, i.e., the DIP, DOP and APIs, is tested in the platform. Furthermore, the development environment is taught as the primary content in the class of Cloud Programming in the undergraduate and graduate programs of the Xi'An Technological University. Some students program with the tool to complete the research projects. Now the project of New World Wide Web has reached the lines of code more than 550,000. It is designed to enclose an unlimited number of computers as clusters to be as scalable as possible to deal with the potentially high volume accessing from users over the globe. Additionally, since the development tool is invented, the implementation efficiency is raised apparently. Because it is necessary to confront the most difficult distributed environment, i.e., the large-scale heterogeneous social distributed computing circumstance [13], it proves further that GreatFree has sufficient flexibility to deal with the extreme case without high implementation overhead.

It is believed that the issue of programming in the diverse distributed systems, especially over the Internet, is tough. For that, it is necessary to move forward based on the current achievement. Since no proper programming languages are suitable to the current complicated distributed computing environments over the Internet, it is feasible to propose a new language that is asynchronous and distributed by nature rather than synchronous and standalone by default. Moreover, different from the existing ones, such as Akka [6] and Erlang [7], the new language intends to avoid the pursuit of a universal abstract model. In contrast, the new language attempts to propose a series of novel keywords and semantics that are transformed from the DIP, DOP and APIs. Finally, the new language does not think it is necessary to make updates to the current syntax. It will conform to the convention of traditional programming languages, like C, C++ and Java, in order to fit most developers' habits.